\documentclass[aps,prl,groupedaddress,showpacs]{revtex4-1}
\usepackage[latin1]{inputenc}
\usepackage[T1]{fontenc}
\usepackage{graphicx}
\usepackage{dcolumn}
\usepackage{bm}
\usepackage{amsmath}
\usepackage{hyperref}

\def\textsum{\begingroup\textstyle\sum\endgroup}

\begin{document}

\preprint{APS/123-QED}

\title{Feedback control of trapped coherent atomic ensembles}

\author{T. Vanderbruggen$^1$}
\altaffiliation[Now at: ]{ICFO - Institut de Ciències Fotòniques, E-08860 Castelldefels (Barcelona), Spain}
\author{R.~Kohlhaas$^1$}
\author{A. Bertoldi$^{1,3}$}
\author{S. Bernon$^1$}
\altaffiliation[Now at: ]{Universit{\"a}t T{\"u}bingen, D-72076 T{\"u}bingen, Germany}
\author{A. Aspect$^1$}
\author{A. Landragin$^2$}
\author{P. Bouyer$^{1,3}$}
\address{	$^1$Laboratoire Charles Fabry, Institut d'Optique, CNRS, Universit{\'e} Paris-Sud \\
			Campus Polytechnique, RD 128, 91127 Palaiseau, France \\
		$^2$LNE-SYRTE, Observatoire de Paris, CNRS and UPMC \\
			61 avenue de l'Observatoire, F-75014 Paris, France \\
		$^3$Laboratoire Photonique, Num{\'e}rique et Nanosciences - LP2N Universit{\'e} Bordeaux - IOGS - CNRS: UMR 5298 \\
			B{\^a}t. A30, 351 cours de la liberation, Talence, France}
			
\date{\today}

\begin{abstract}

We demonstrate how to use feedback to control the internal states of trapped coherent ensembles of two-level atoms, and to 
protect a superposition state against the decoherence induced by a collective noise. Our feedback scheme is based on weak
optical measurements with negligible back-action and coherent microwave manipulations. The efficiency of the feedback system
is studied for a simple binary noise model and characterized in terms of the trade-off between information retrieval and
destructivity from the optical probe. We also demonstrate the correction of more general types of collective noise. This
technique can be used for the operation of atomic interferometers beyond the standard Ramsey scheme, opening the way towards
improved atomic sensors.

\end{abstract}


\maketitle

Coherent ensembles of two-level atoms, whose quantum evolution provides a fundamental oscillatory signal, are the core of
many instruments based on Ramsey interferometry, like atomic clocks, magnetic field sensors, or inertial and gravitational
sensors \cite{cronin2009,budker2007,guena2012}. Trapping such atomic
ensembles gives access to long observation times, and thus extreme precision, provided that one can fight the loss of
coherence induced by ambient noise. A noise homogeneous over the size of the ensemble
affects all the atoms in the same way. If the number of atoms is large, it becomes possible to measure the effect of this
collective noise with negligible perturbation of the state of the individual systems. This can be done using weak
measurements, as proposed in \cite{lloyd2000}, and used for example to determine the collective atomic state of a coherent
ensemble of atoms \cite{smith2006}. It is then possible to react on the atoms to compensate for the effect of the noise, and
thus fight the corresponding decoherence. Similar techniques have been proposed to boost the performance of atomic clocks by
phase-locking the local oscillator to the atomic phase \cite{shiga2012}.

In this Letter, we demonstrate such a measurement and correction scheme, using a trapped coherent ensemble of two-level
atoms. The atomic system consists in rubidium 87 atoms prepared in a coherent superposition of the two ground hyperfine
levels $\left| \, 0 \, \right\rangle \equiv \left| \, F=1,m_{F}=0 \, \right\rangle$ and $\left| \, 1 \, \right\rangle \equiv
\left| \, F=2,m_{F}=0 \, \right\rangle$ of the electronic ground state ${5\, ^{2}}$S$_{1/2}$. This superposition can be
manipulated by the interaction with a microwave field resonant with the 6.835 GHz transition between the two levels. The
population difference is weakly measured using a frequency modulated optical probe, where the sidebands are phase shifted
with opposite signs by the two atomic populations. This probe is used to evaluate the effect of a collective noise on the
atoms for later correction. We consider two different noise models: the first model takes randomly one of two known values,
and the second one takes a random value uniformly distributed.

A sample of $N_{\rm at}$ indistinguishable two-level atoms can be represented as an ensemble of effective spins 1/2, whose
sum defines a collective spin, or Bloch vector, with observables $\mathbf{J} = \left( J_{x}, J_{y}, J_{z} \right)$ on the
Bloch sphere \cite{arecchi72,itano93}. The observable $J_{z}$ refers to the population difference between the two atomic
levels, while $J_{x}$ and $J_{y}$ characterize the coherence between the two levels. When all the atoms are in the same pure
single particle state, they form a coherent spin state (CSS) or Bloch state; the associated Bloch vector has its extremity on
a Bloch sphere of radius $J=N_{\rm at}/2$. Any collective and homogeneous interaction with the microwave results in a
rotation of the Bloch vector. We restrict our study to rotations around $Y$: all accessible states are then represented by a
vector in the $y=0$ plane, labelled $\left| \, \theta \, \right\rangle$ from the angle $\theta$ it forms with the $Z$ axis.

\begin{figure}[b]
\centering
\includegraphics[width=12cm]{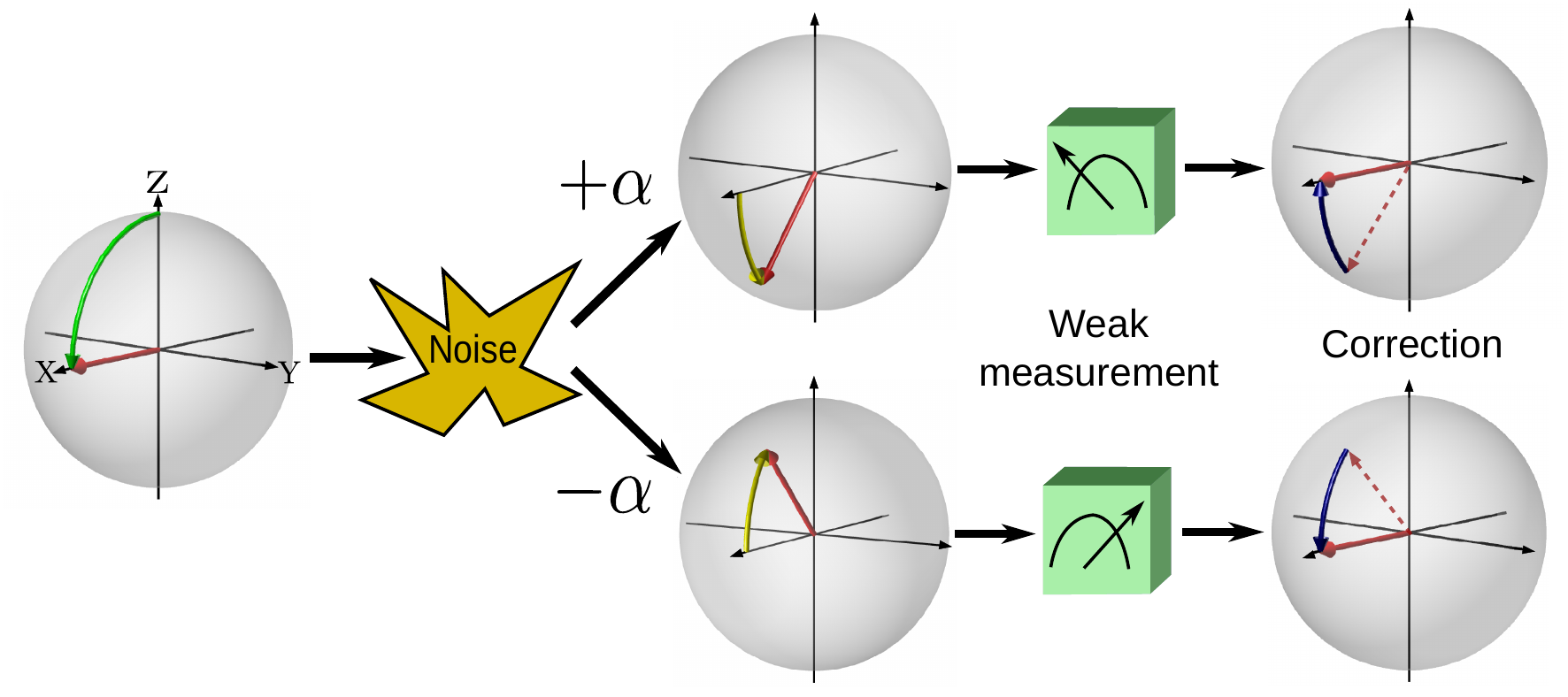}
\caption{\label{fig:scheme} (color online). Evolution of the collective spin on the Bloch sphere (case of a binary random
collective rotation). A $\pi / 2$ rotation around $Y$ prepares a coherent superposition. The state experiences a random
rotation of $+\alpha$ or $-\alpha$ around $Y$, which is detected using a weak non-destructive measurement and then
corrected.}
\end{figure}

In our experiment we initially prepare the atoms in the CSS $\left| \, \theta = \pi / 2 \, \right\rangle$, which is the state
after the first beamsplitter in a Ramsey interferometer sequence and for which $\left\langle J_z \right\rangle=0$. Our goal
is to recover that state after it is submitted to a collective noise (Fig. \ref{fig:scheme}). The noise consists of random
collective rotations (RCRs) implemented using microwave pulses that rotate the Bloch vector around the $Y$ axis. The rotation
angle is selected randomly according to each specific model. The RCR transforms the initial state into a statistical mixture
of all the states that can be generated by the noise; as a consequence the length of the Bloch vector decreases from
$J=N_{\rm at}/2$ to a lesser value, thus decreasing the coherence of the initial state. We have implemented two noise models:
first, a binary RCR, where the collective Bloch vector is submitted to rotations picked randomly among two fixed values
$+\alpha$ and $-\alpha$; second, an analog RCR with a random rotation angle uniformly distributed between $-\alpha$ and
$+\alpha$. The binary RCR transforms the initial CSS into a balanced statistical mixture of the states $\left| \, {\pi /
2}+\alpha \, \right\rangle$ and $\left| \, {\pi / 2}-\alpha \, \right\rangle$. The analog RCR transforms the initial CSS into
the statistical mixture $\rho = 1/(2\alpha) \int_{-\alpha}^{+\alpha} \left| \, {\pi / 2}+\theta \, \right\rangle \left\langle
{\pi / 2}+\theta \right| d \theta$. The coherence of the atomic ensemble then decreases from unity to  $\eta_{\alpha} = \cos
\alpha$ in the binary case, and to $\eta_{\alpha} = {\sin \alpha / \alpha}$ in the analog case.

To fight the loss of coherence induced by the noise, we optically measure $J_{z}$ after each random rotation, and apply a
counter-rotation depending on the result. We consider in this work only measurements with an uncertainty $\sigma$ larger than
the atomic projection noise ($\sigma \gg \sqrt{N_{\rm at}}$). For large CSSs, the measurement uncertainty is small compared
to the length of the spin ($\sigma < J$) and a weak measurement can lead to precise information with negligible back-action.
In the binary RCR case, the probability to detect the hemisphere in which the Bloch vector lies ($z$>0 or $z$<0) is
\begin{equation}
p_s = \int_0^{\infty} P(m_0|-\alpha) \, \mathrm{d} m_0 = \frac{1}{2} \left[ 1 + \mathrm{erf} \left(\frac{\sqrt{2} \; J\sin{\alpha}}{\sigma} \right) \right],
\label{eq:erf}
\end{equation}
where $P(m_0|-\alpha)$ is the probability to obtain $m_{0}$ when measuring $J_{z}$ given a noise rotation $-\alpha$. After
the correction the system is in a statistical mixture of the initial state $\left| \, {\pi / 2} \, \right\rangle$, recovered
with probability $p_{s}$, and of the two states $\left| \, {\pi / 2} \pm 2\alpha \, \right\rangle$ resulting from a wrong
estimation that doubles the RCR angle. The coherence of this statistical mixture is
\begin{equation}
\eta_{\alpha}^{\rm out} = \left[ p_{s} + (1-p_{s}) \cos (2 \alpha) \right] \; e^{-\gamma N_{\rm ph}} \; ,
\label{eq:cohBinary}
\end{equation}
where the exponential factor accounts for the effect of the spontaneous emission induced by $N_{\rm ph}$ photons
in the probe pulse, and the coefficient $\gamma$, which depends on the resonant optical density, is determined
experimentally (see below).

\begin{figure}
\centering
\includegraphics[width=12cm]{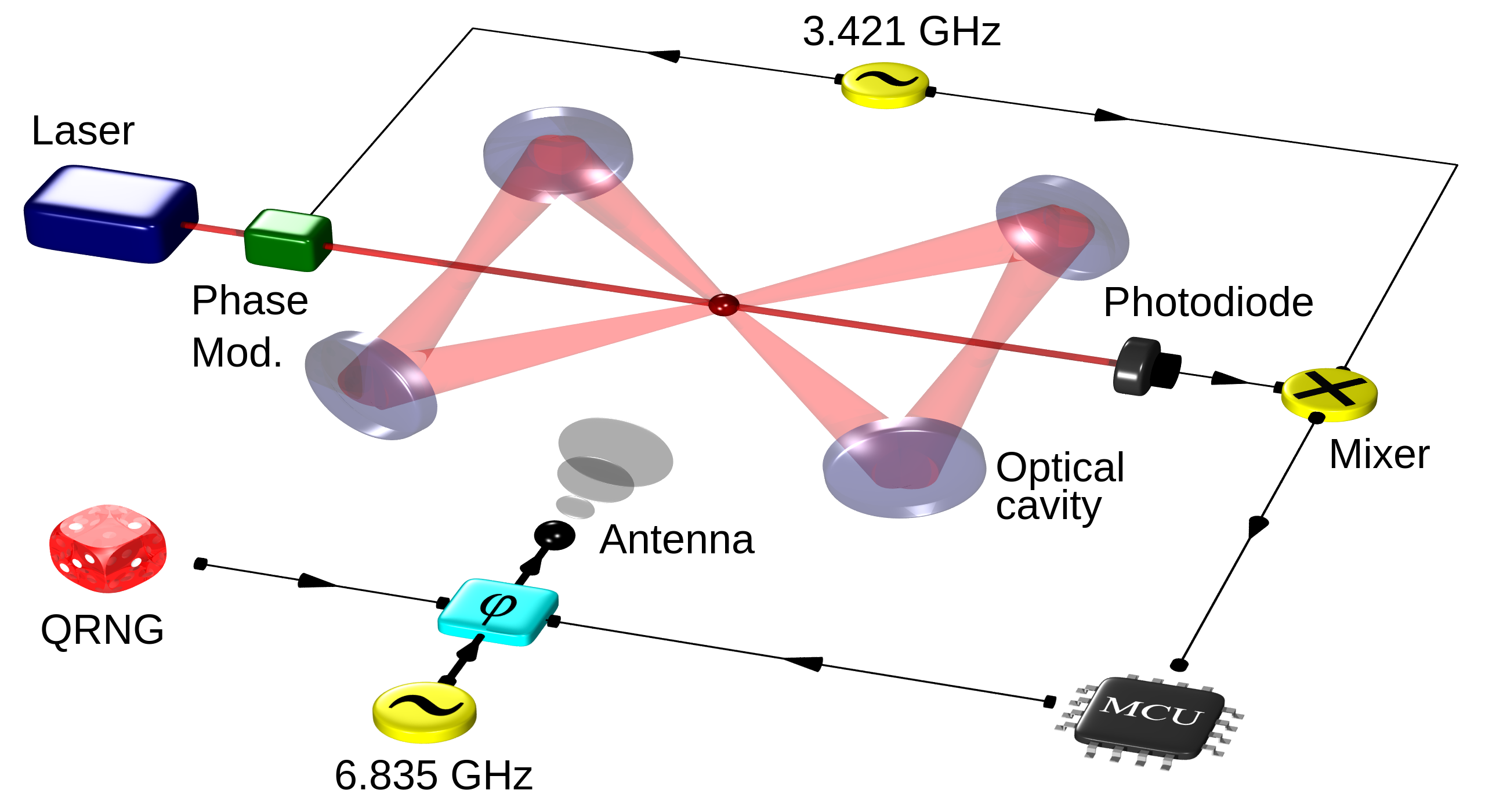}
\caption{\label{fig:setup} (color online). Experimental setup: the atomic sample is coherently manipulated with the microwave
field emitted by an antenna. The detection laser at 780 nm is phase modulated at 3.421 GHz before passing through the atomic
cloud and being detected on a photodiode. After demodulation, the signal is digitized and sent to a microcontroller unit that
computes and sets the phase of the microwave. A quantum random number generator (QRNG), connected to the phase-shifter
($\varphi$), implements the RCR.}
\end{figure}

We work with $5\times10^{5}$ $^{87}$Rb atoms at T=10 $\mu$K, optically trapped by laser light at 1550 nm (Fig.
\ref{fig:setup}). The laser intensity is enhanced using a 4 mirror optical resonator \cite{bernon2011}. The atomic cloud
(radius at 1/$e^2$ of 50 $\mu$m) is trapped where two cavity arms (waist of 100 $\mu$m) cross. The radiation that traps the
atoms in $| \, 0 \, \rangle$ and $| \, 1 \, \rangle$, generates a strong, spatially inhomogeneous broadening of the D$_2$
transition \cite{bertoldi2010}, used for the probing. The effect is compensated by modifying the light shift of the $5^2{\rm
P}_{3/2}$ level using an auxiliary, spatially matched laser beam \cite{supplMat}. The non-destructive detection of $J_{z}$ is
based on the phase-shift that the atomic sample induces on a far off-resonance optical probe
\cite{bernon2011,appel08,schleier-smith10,koschorreck2010,kohnen2011,vanderbruggen2011}. The probe beam has a waist of 245
$\mu$m on the atomic sample. It is phase modulated at a frequency $\Omega=3.421$ GHz, and frequency referenced so that each
sideband mainly probes the population of one of the two levels $\left| \, 0 \, \right\rangle$ and $\left| \, 1 \,
\right\rangle$, with the same magnitude and opposite sign for the couplings \cite{supplMat,saffman2009}. We cancel the probe
induced light shift and the related decoherence by precisely compensating the effect of the carrier with that of the
sidebands \cite{supplMat}.

We prepare the initial CSS $\left| \, \theta = \pi / 2 \, \right\rangle$ by optically pumping the atoms in $\left| \, 0 \,
\right\rangle$ and applying a ${\pi / 2}$ microwave pulse of duration $\tau_{\pi / 2} = 75.6(2)~\mu$s. We study the control
process consisting of a binary RCR, a weak measurement pulse to determine the sign of $J_z$, and a correcting rotation. The
binary RCR is implemented as a $\alpha = {\pi / 4}$ microwave pulse; the coherence after the noise and after the correction
is $\eta_{\pi / 4}=1/\sqrt{2}$ and $\eta_{\pi / 4}^{\rm out}=p_s \; e^{-\gamma N_{\rm ph}}$ (see Eq. (\ref{eq:cohBinary})),
respectively. The spontaneous emission due to the carrier is negligible compared to that of the sidebands, hence $N_{\rm ph}$
relates to the number of photons in each sideband. The rotation sign is randomly chosen by a quantum random number generator
(Quantis, IDQuantique) and set through a phase-shifter on the microwave. The $J_{z}$ measurement uses a 1.5~$\mu$s long probe
pulse. The detected signal is demodulated to get the population difference, and then  analogically integrated to obtain its
mean value over the pulse length. To implement the feedback, the output of the integrator is digitized and treated in
real-time with a microcontroller to get the sign of $J_{z}$. The latter controls the rotation direction for the ${\pi / 4}$
correction pulse through the microwave phase-shifter.

\begin{figure}
\centering
\includegraphics[width=12cm]{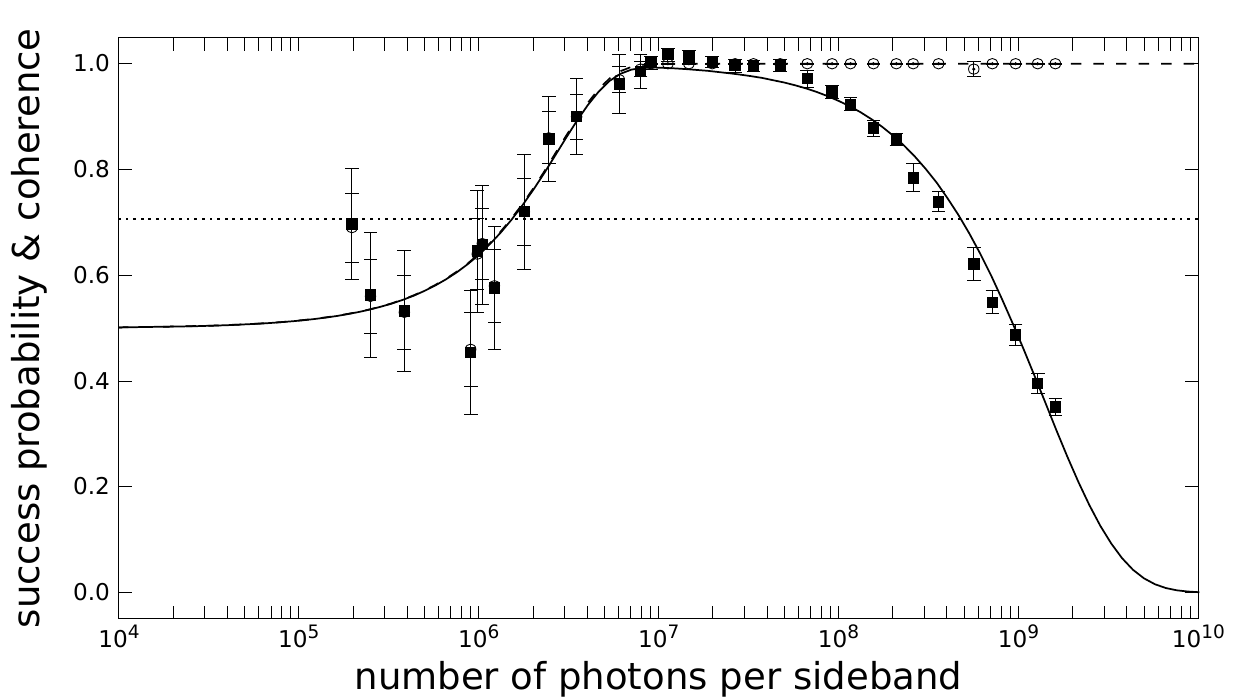}
\caption{\label{fig:feedback_res_coherence} Remaining coherence (solid squares) after the cycle consisting of a binary RCR,
measurement and correction; success probability (open circles) versus the number of photons per sideband in the probe pulse.
The solid line is a fit of the remaining coherence data with Eq. (\ref{eq:cohBinary}), the dashed one of the success
probability data with Eq. (\ref{eq:erf}). The dotted line at 1/$\sqrt{2}$ indicates the coherence after the RCR. Error bars
are the $\pm1$ standard error of statistical fluctuations.}
\end{figure}

We evaluate the efficiency of the feedback control in terms of the achieved coherence recovery. To determine the coherence of
the atomic state at the end of the cycle, we send a second $\pi / 2$ pulse to obtain a Ramsey type measurement. Fig.
\ref{fig:feedback_res_coherence} shows the remaining coherence after one cycle (solid squares), when the measurement
uncertainty $\sigma$ is varied by changing the number of photons in the probe pulse. Each point results from 50 repetitions
of the sequence, and the reported error bars reflect the statistical spread. We fitted the data set with Eq.
(\ref{eq:cohBinary}) for $\eta_{\pi / 4}^{\rm out}$, using Eq. (\ref{eq:erf}) for the success probability $p_{s}$ and
assuming for $\sigma$ two contributions, one related to the photonic shot noise ($\propto {N_{\rm ph}}^{-1/2}$) and one to
technical noise ($\propto {N_{\rm ph}}^{-1}$): the latter is dominant and we obtain $\sigma = 9.6(5)\times 10^{11}/N_{\rm
ph}$ \cite{note2}. The fit of Fig.~\ref{fig:feedback_res_coherence} yields also the rate for the probe induced decoherence
per photon $\gamma=7.6(4) \times 10^{-10}$. The remaining coherence of the output state reaches an optimum of 0.993(1) with
$9.1 \times 10^{6}$ photons per sideband: this value exceeds the coherence $1/\sqrt{2} \approx 0.707$ of the mixed state
after the RCR, proving the efficiency of our scheme.

We confirmed this result by multiplying the success probability $p_s$ of detecting the right hemisphere and the probe induced
decoherence measured separately using Ramsey interferometry. To obtain $p_s$, the sign of the RCR and the corresponding
correction are recorded during the experiment; treated off-line they produce the open circles of Fig.
\ref{fig:feedback_res_coherence}, fitted with Eq. (\ref{eq:erf}).

To study how the feedback scheme can protect a CSS over time in the presence of noise, we iterate 200 times on the same
atomic ensemble the cycle consisting of the binary RCR of angle $\pi / 4$, the weak measurement of $J_z$, and the
corresponding correction rotation. At each cycle $J_{z}$ is measured by integrating the signal determined by $1.4 \times
10^{7}$ photons in each sideband.

\begin{figure} \centering \includegraphics[width=12cm]{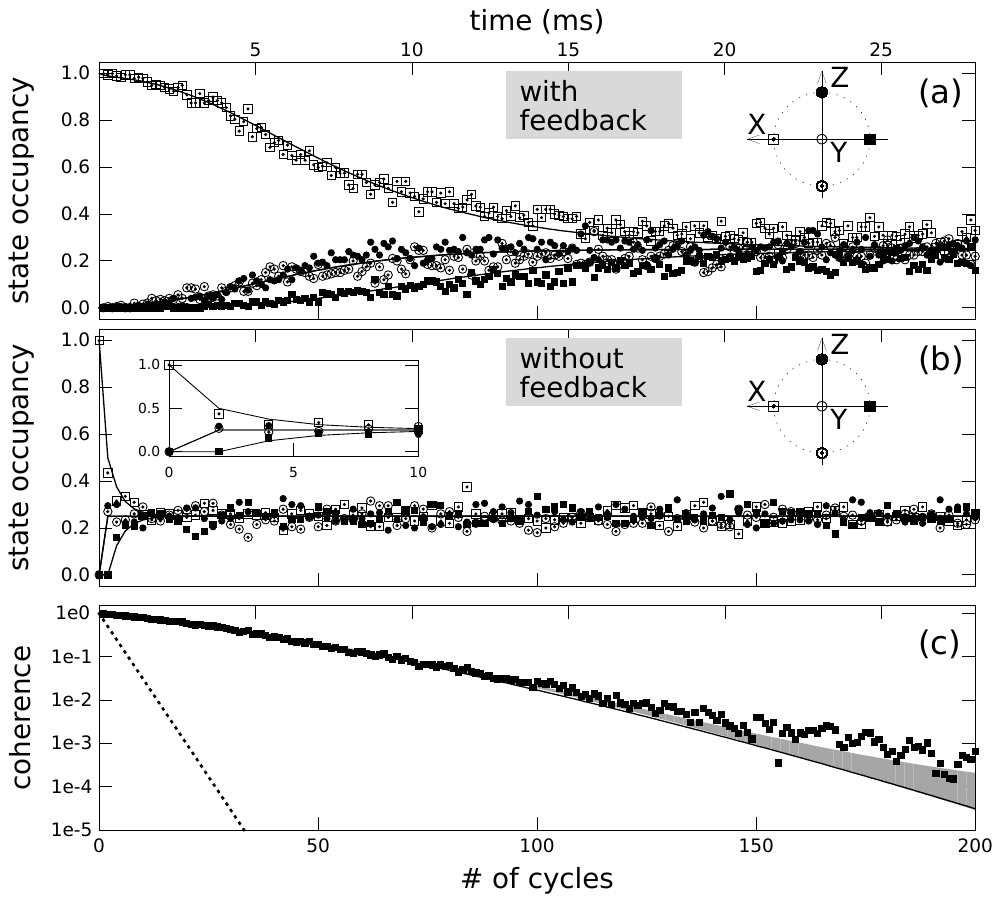}
\caption{\label{fig:coherence_seq} State occupancy versus the number of cycles for the state $\left| \, 0 \, \right\rangle$
(solid circles), $\left| \, \pi / 2 \, \right\rangle$ (open squares), $\left| \, \pi \, \right\rangle$ (open circles) and
$\left| \, 3\pi / 2 \, \right\rangle$ (solid squares) with (a) and without (b) feedback correction. Each experimental point
is obtained from 200 repetitions of the sequence. One cycle lasts 140 $\mu$s. The solid lines are calculated independently,
considering the probabilistic outcome of the RCRs, and in closed-loop that of the corresponding corrections. Inset: In the
open loop case, the state occupancy is equally distributed over the four states after about 10 cycles. (c) Calculated
remaining coherence with (solid line) and without (dashed line) feedback. The experimental points in closed-loop are shifted
at the top of the shaded region because of the finite size statistical sample in the state occupancy determination.}
\end{figure}

During that experiment the signs of the RCRs and the corresponding corrections are recorded; analyzed off-line they provide
the trajectory followed by the Bloch vector. The state occupancy, which is the probability to be in a given state, is
measured versus time averaging the results of 200 experimental runs. In the closed loop case, the system spreads from $\left|
\, \pi / 2 \, \right\rangle$ to the two poles ($\left| \, 0 \, \right\rangle, \left| \, \pi \, \right\rangle$), and at a
slower rate to $\left| \, 3 \pi / 2 \, \right\rangle$ (Fig.~\ref{fig:coherence_seq}(a), points). The evolution of the state
occupancy over the four states is explained in terms of the random rotations of $\pm {\pi / 4}$ at every cycle, and by the
success probability $p_s$ of the weak measurement (Fig.~\ref{fig:coherence_seq}(a), solid lines). At each cycle, $p_s$
decreases since the spontaneous emission induced by the probe and the residual inhomogeneous differential light shift of the
trap shorten the spin. The trap induced decoherence has been characterized by Ramsey interferometry, and the coherence loss
versus the time in the dipole trap shows a Gaussian decay. Considering the cycle duration, the trap decoherence can be
expressed in terms of the number of cycles $N$ as $\mathrm{exp}(-\left(N/N_0\right)^2 )$, where $N_0=157.6$. This decoherence
is not a limitation in the present case and could be further reduced using a compensation laser beam
\cite{kaplan2002,radnaev2010}. 

The state occupancy is compared to the open-loop case, where measurements and corrections are not applied: after a few
iterations the state vector reaches a balanced statistical mixture of four states: $\left\{ \left| \, 0 \, \right\rangle,
\left| \, \pi / 2 \, \right\rangle, \left| \, \pi \, \right \rangle, \left| \, 3 \pi / 2 \, \right\rangle \right\}$ for an
even number of iterations, and $\left\{ \left| \, \pi / 4 \, \right\rangle, \left| \, 3 \pi / 4 \, \right \rangle, \left| \,
5 \pi / 4 \, \right \rangle, \left| \, 7 \pi / 4 \, \right \rangle \right \}$ for an odd number
(Fig.~\ref{fig:coherence_seq}(b), points). Here the state occupancy evolution results only from the random rotations of $\pm
{\pi / 4}$ at every cycle (Fig.~\ref{fig:coherence_seq}(b), solid lines).

The remaining coherence has been evaluated by multiplying the effects of the decoherence sources (Fig.
\ref{fig:coherence_seq}(c)), while the spin rotations are considered perfect. In open-loop the remaining coherence given by
the spin diffusion, equal to ${{\eta_{\pi / 4}}^N=1/2^{N/2}}$, has to be multiplied by the effect of the trap. In closed-loop
the factors to consider to obtain the coherence reduction are given by the state occupancy evolution, the probe spontaneous
emission, and the trap. The feedback correction greatly improves the coherence lifetime of the system even when the
experimental imperfections and limitations are taken into account: for example, after $N=10$ cycles the remaining coherence
without feedback is 0.03 whereas it reaches 0.77 when a correction is applied.

We now study the case of an analog RCR with a random rotation angle uniformly distributed on $[-{\pi / 2},+{\pi / 2}]$ and
acting on $\left| \, \pi / 2 \, \right\rangle$; this generates a statistical mixture with coherence $\eta_{\pi / 2}=2/\pi$.
The analog RCR is implemented using the quantum random number generator to control both the length of the microwave pulse and
the rotation sign. After the RCR, a probe pulse with $2.8 \times 10^{7}$ photons is sent to measure $J_{z}$. From the
measurement result, we set both the length and the direction of the correction pulse. The feedback bandwidth is about 10 kHz
and is limited by the duration of the correction pulse, which is set as a function of the measured value of $J_{z}$.

As reported for the binary RCR case, the coherence of the state after one cycle with and without feedback is directly
measured by Ramsey interferometry. Averaging over 400 repetitions, the coherence obtained without feedback is $0.63(3)$,
consistent with the expected value of $2/\pi \approx 0.637$; the correction pulse increases the coherence to $0.964(5)$.

We can again compare the measured value of the coherence after one cycle to the value obtained by multiplying the effects of
the different decoherence sources. To evaluate the decoherence due to the direction spread of the spin, during each sequence
we record the length $\tau_{N}$ and the direction $\epsilon_{N}$ (+/-1 for a pos./neg. rotation) for the noise pulse, and the
corresponding parameters $\tau_{C}$ and $\epsilon_{C}$ for the correction pulse. The spin direction is then computed as
$\theta_{N} =\pi \epsilon_{N}( \tau_{N}/ \tau_{\pi })$ for the CSS after the noise pulse, $\theta_{C} = \theta_{N}+\pi
\epsilon_{C}( \tau_{C} /\tau_{\pi })$ after the correction pulse. For 5000 repetitions of the sequence $\theta_{N}$ is
uniformly distributed over $[-{\pi / 2},+{\pi / 2}]$ whereas $\theta_{C}$ is well described by a Gaussian distribution
centered at zero and of standard deviation 207(10) mrad. The angular spread is explained in terms of the measurement
uncertainty of 6.8\% over $J_z$, increased by a factor 2 because of a resolution loss in the digital controller. The
remaining coherence for the spin spread is 0.979(2). The probe induced spontaneous emission reduces the coherence by another
factor 0.979(1). The product of these two factors gives a final remaining coherence of 0.958(2), which is consistent with its
direct measurement.

To summarize, we have demonstrated the partial protection of an atomic CSS from the decoherence induced by RCRs around a
fixed axis, using feedback control based on weak non-destructive measurements. The method can be generalized to rotations
around an arbitrary axis of the Bloch sphere: one could consecutively read out $J_x$, $J_y$ and $J_z$ and correct with
suitable rotations. Compared to spin-echo techniques, relying on temporal invariance of the noise, our feedback method allows
the compensation of time dependent noise, provided that the time evolution is slower than the correction time. By increasing
the effective on-resonance optical depth, the feedback scheme could be implemented in the projective limit to
deterministically prepare non-classical states \cite{wiseman94,thomsen02,stockton2004}, using measurement based spin
squeezing \cite{appel08,schleier-smith10,chen2011,sewell2012}. Finally, we note that coherence preserving techniques that
combine repeated measurements and feedback pave the way towards novel atom interferometry schemes, as recently proposed with
atomic clocks to achieve white phase noise \cite{shiga2012}.

We thank A. Browaeys for comments. We acknowledge funding from DGA, IFRAF, CNES, the European Union (EU) (iSENSE), EURAMET
(QESOCAS), ANR (MINIATOM), and ESF Euroquam. A.~A. acknowledges support from ERC QUANTATOP, P.~B. from a chair of excellence
of R{\'e}gion Aquitaine.

\newpage

\section{SUPPLEMENTAL MATERIAL}

\begin{figure}[!h]
\centering
\includegraphics[width=12cm]{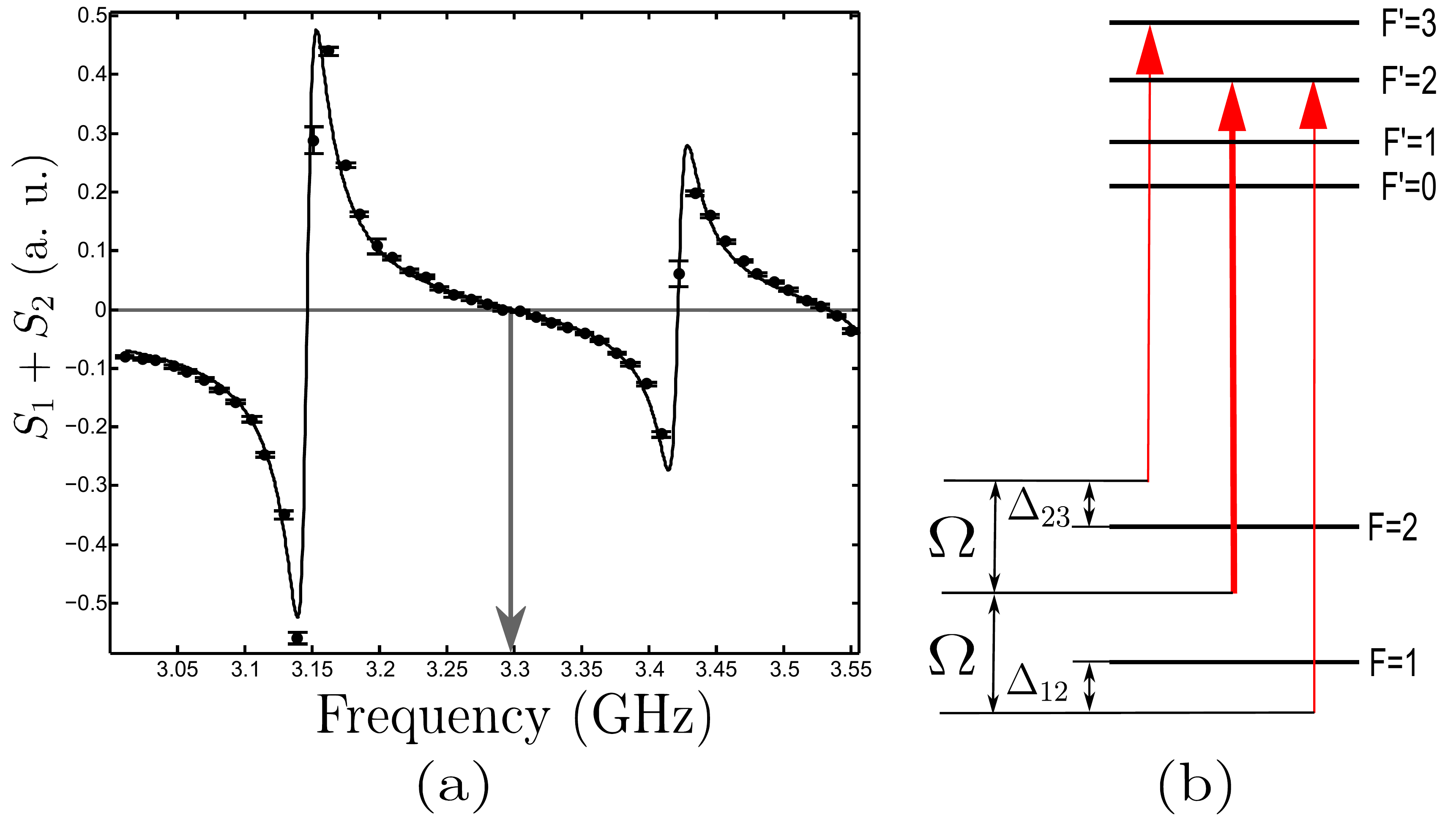}
\caption{\label{fig:Jz_adjustement} (a) Coupling of the probe versus the position of the carrier with respect to the $\left|
F=1 \right\rangle \rightarrow \left| F'=2 \right\rangle$ transition. The experimental results (filled circles) are compared to
the calculated coupling (solid line). (b) Hyperfine structure of the D$_{2}$ transition of $^{87}$Rb and relative position of
the probing beams. The thick line is the carrier whereas the thin lines are the sidebands.}
\end{figure}

\noindent \textbf{Measurement of the observable $J_{z}$.} An optical beam is phase modulated to have 4.6\% of the total power
in each sideband and passes through the atomic sample before being detected on a fast photodiode. With the two sidebands
generated by the phase modulator it is possible to directly measure the population difference between the hyperfine levels of
the ground state for a $^{87}$Rb atomic ensemble: one sideband is placed close to the $\left| F=1 \right\rangle \rightarrow
\left| F'=2 \right\rangle$ transition, the other one close to the $\left| F=2 \right\rangle \rightarrow \left| F'=3
\right\rangle$ transition, as depicted in Fig.~\ref{fig:Jz_adjustement}(b). The coupling $S_{1}$ ($S_{2}$) of the first
(second) sideband with the $\left| F=1 \right\rangle \rightarrow \left| F'=2 \right\rangle$ ($\left| F=2 \right\rangle
\rightarrow \left| F'=3 \right\rangle$) transition, is given by:
\begin{equation}
S_{F} = \textsum_{F'} \frac{\gamma \Delta_{FF'}}{\Delta_{FF'}^{2} + \gamma^{2} \left( 1 + I/I_{\rm sat} \right)} S_{FF'},
\label{eq:coupling_SF}
\end{equation}
where $\gamma$ is the half width at half maximum of the atomic transition, $I$ the intensity of each sideband, $I_{\rm sat}$
the saturation intensity of the transition and $S_{FF'}$ the dipolar coupling associated to the $\left| F \right\rangle
\rightarrow \left| F' \right\rangle$ transition \cite{steck}. The phase-shift induced by the atomic sample on the probe is
proportional to $ \phi_{\rm at} \propto N_{1} S_{1} + N_{2} S_{2}$, where $N_{1}$ ($N_{2}$) is the population in the $\left|
F=1 \right\rangle$ ($\left| F=2 \right\rangle$) level. If the detunings $\Delta_{FF'}$ are adjusted so that $S_{1} = -
S_{2}$, then $\phi_{\rm at} \propto N_{1}- N_{2}$ and the detection measures the observable $J_{z}$.

To adjust the detunings $\Delta_{FF'}$, we first set a modulation frequency $\Omega = 3.421$ GHz and prepare the atoms in a
coherent superposition $\left| \pi / 2 \right\rangle$ so that $N_{1} = N_{2}$. We measure then the demodulated signal versus
the position of the carrier with respect to the $\left| F=1 \right\rangle \rightarrow \left| F'=2 \right\rangle$ transition.
The result in Fig.~\ref{fig:Jz_adjustement}(a), in very good agreement with Eq.~(\ref{eq:coupling_SF}), was obtained with a
carrier power of 153 $\mu$W and a power per sideband of 7.1 $\mu$W. The beam waist of the probe on the atomic sample is 245
$\mu$m, which gives an intensity on the sample of 11.2 mW/cm$^{2}$. Since a $\pi$ transition is probed, the saturation
intensity is $I_{\rm sat} = 2.503$ mW/cm$^{2}$ \cite{steck}. The condition $S_{1} + S_{2} = 0$ is reached when the carrier
is set at 3.291 GHz from the $\left| F=1 \right\rangle \rightarrow \left| F'=2 \right\rangle$ transition, and the
detunings of the sidebands are $\Delta_{12} =$ -126.7 MHz and $\Delta_{23} =$ 148.5 MHz. \\

\begin{figure}[!ht]
\centering
\includegraphics[width=12cm]{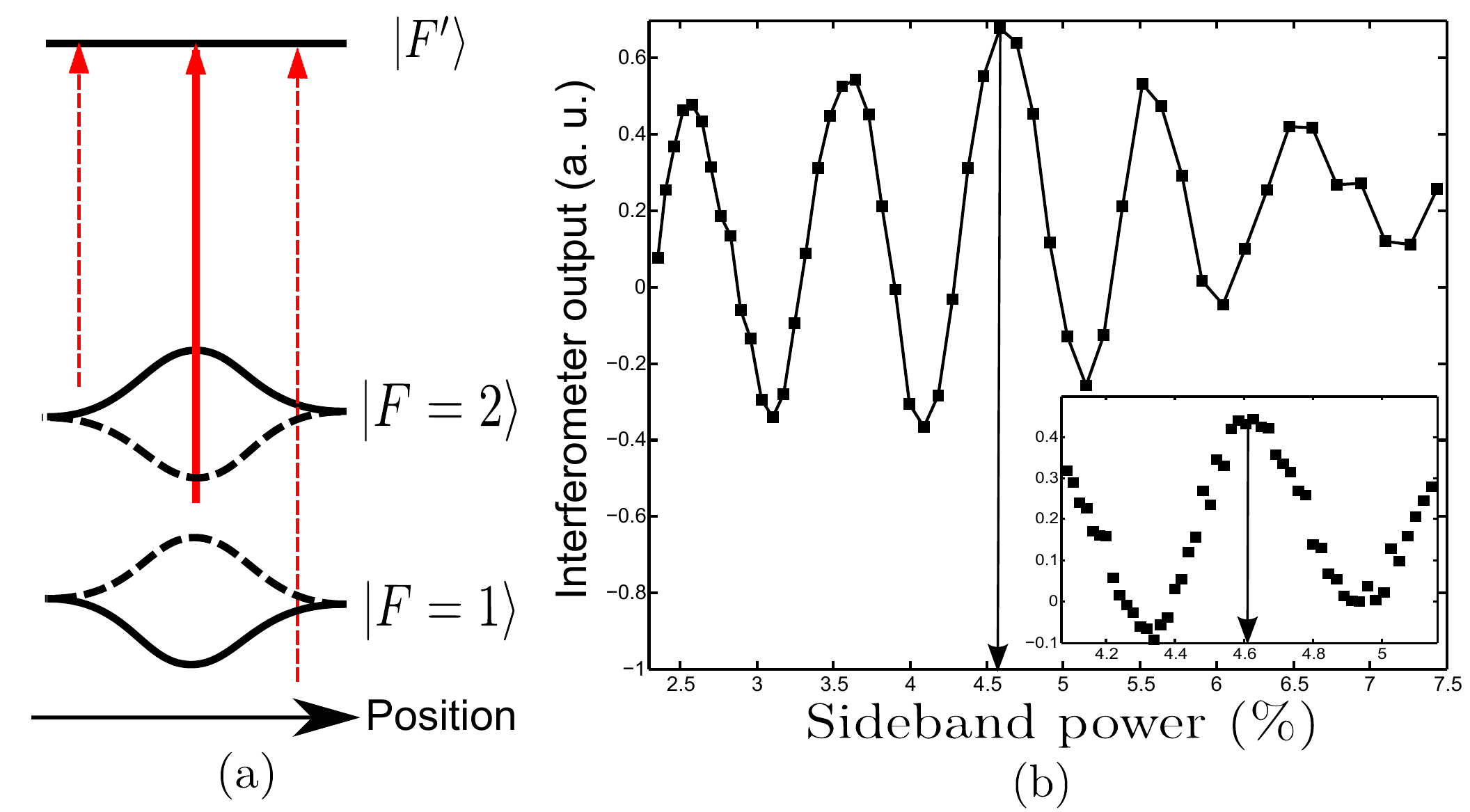}
\caption{\label{fig:light_shift_comp_sidebands} (a) Light shift induced by the carrier (solid lines) and by the sidebands
(dashed lines) versus the position of an atom in the probe beam. (b) Output of a Ramsey interferometer versus the power in
each sideband in percent of the carrier power. A 40~$\mu$s probe pulse is sent in between the two $\pi / 2$ microwave
pulses. In inset, the same measurement is performed with a 70~$\mu$s long pulse to determine more precisely the compensation
ratio.}
\end{figure}

\noindent \textbf{Compensation of the probe induced light shift.} The light shift of the probe is a severe limitation in most
applications using non-destructive probing techniques, especially for interferometry because it dephases the atomic sample.
Here, it induces not only a rotation of the mean spin around the $Z$ axis of the Bloch sphere but also spreads the spins due
to the spatial inhomogeneity of the Gaussian probe beam. This spread acts as a decoherence source and prevents to realize the
feedback scheme if not compensated. The symmetry of the coupling of each sideband with the related transition allows us to use
the light shift generated by the carrier to equally compensate that of each sideband
(Fig.~\ref{fig:light_shift_comp_sidebands}(a)). Moreover, since the carrier and the sidebands are spatially overlapped the
compensation is homogeneous. The compensation is realized by adjusting the relative power between the sidebands and the
carrier. The tuning of the power ratio is precisely set using a Ramsey sequence with a probe pulse sent between the two $\pi /
2$ pulses. Interference fringes are scanned by changing the intensity in the sidebands
(Fig.~\ref{fig:light_shift_comp_sidebands}(b)) and the condition of maximum contrast gives the position of the zero light
shift; this is obtained with $4.6$\% of the total power in each sideband.

\noindent \textbf{Compensation of the differential light shift on the D$_{\mathbf 2}$ line.} Due to the $5^{2}P_{3/2}
\rightarrow 4^{2}D_{5/2,3/2}$ transitions at 1529.3 nm \cite{lee2007}, the trapping radiation at 1550 nm induces a
differential light shift on the D$_{2}$ transition \cite{bertoldi2010} used for the non-destructive probing. At the center of
the trap, where the radiation intensity is maximal, the differential frequency shift is about 100~MHz for the typical power in
the optical trap: this is of the same order of $\Delta_{12}$ and $\Delta_{23}$. In such conditions, the coupling of each atom
with the probe is inhomogeneous and this hampers the measurement of the observable $J_{z}$. To compensate for the differential
light shift, a radiation beam at 1528.7 nm, that is about 0.6 nm on the blue side of the $5^{2}P_{3/2} \rightarrow
4^{2}D_{5/2,3/2}$ transitions, is injected in the fundamental mode of the cavity \cite{kohlhaas2012}. By adjusting the power
ratio between the 1550 nm and the 1529 nm beams, we compensate for the differential light shift with a high spatial
homogeneity thanks to the good overlap between the fundamental modes of the cavity at 1550 nm and 1529 nm. This compensation
allows for a quasi-homogeneous coupling of the non-destructive probe with the atomic sample over the spatial extension of the
dipole trap. The residual light shift due to the non perfect mode overlap is about 0.7 MHz, which is not only negligible
compared to $\Delta_{12}$ and $\Delta_{23}$ but also small compared to the transition linewidth ($\Gamma \sim 2 \pi \cdot 6$
MHz).

%\bibliographystyle{plain}	% (uses file "plain.bst")
%\bibliography{sm}		% expects file "myrefs.bib"

\begin{thebibliography}{22}
\bibitem[]{note2}
  \bibinfo{note}{A consistent uncertainty value of $\sigma = 3.4\times10^{4}$ was obtained by measuring the operator $J_z$ on
  $5\times10^{5}$ atoms in the $\left| \, \pi / 2 \, \right\rangle$ state with $N_{\rm ph}=2.8\times10^{7}$. In the same
  conditions the width of the atomic distribution is $\sigma_{\rm at}=7.1\times10^{2} \ll \sigma$, which proves the weakness
  of the measurement}.
\end{thebibliography}

\end{document}